\def\beg{\begin{equation}}
\def\eeq{\end{equation}}
\documentstyle[12pt]{article}
\begin{document}
\begin{center}
{\Large{\bf New resonances along with cyclotron 
resonance in heterostructures: A new radiation.}}
\vskip0.35cm
{\bf Keshav N. Shrivastava}
\vskip0.25cm
{\it School of Physics, University of Hyderabad,\\
Hyderabad  500046, India}
\end{center}

The usual cyclotron resonance occurs at 
$\omega_c=eB/m^*c$. The new resonances occur at 
$\omega_{c\pm}={1\over 2}g_{\pm}eB/m^*c$ where 
${1\over 2}g_{\pm}=({\it l}+{1\over 2}\pm s)/(2{\it l} 
+1)$. The
 energy in the centre of two resonance frequencies 
 varies as the 
square root of the two-dimensional density of the
 electrons due to 
spin access in the Gaussian model. The frequencies
 $\omega_{c\pm}$
 are linearly proportional to the magnetic field 
except near crossing
 point where the linear combination of wave functions
 must be made, i.e., ${1\over \sqrt 2}(|n,{\it l},
 \uparrow>\pm|n, {\it l}, \downarrow>)$.
\vskip0.10cm
\noindent Corresponding author: keshav@mailaps.org\\
Fax: +91-40-2301 0145.Phone: 2301 0811.
\vskip1.0cm

\noindent {\bf 1.~ Introduction}

     When an external magnetic field is applied to 
the electrons they
 go into cyclotron orbits. When energy is swept, there
 is a  resonance at $\omega_c=eB/m^*c$. Since $e$ and
 $c$ are already known and $B$ 
can be measured accurately, a measurement of 
$\omega_c$ leads to a measurement of the effective 
mass of the electron. Here $\omega_c$ 
is called the cyclotron frequency, $e$ the charge of 
the electron, 
$c$ the velocity of light, $m^*$ the effective mass of
 the electron 
and $B$ the external magnetic field[1].

     In this paper, we propose that there must exist
 new resonances 
at $\omega_{c+}={1\over 2}g_+eB/m^*c$ and at 
$\omega_{c-}={1\over 2}
g_-eB/m^*c$. The $g_{\pm}$ are related by Kramers time
 reversed states and the energy at the centre of these
 states is proportional to the square root of the 
two-dimensional number density of electrons. We 
have found[2-7] the factor before the charge during
 our study of the quantum Hall effect where it is used 
to describe the effective fractional charge of the
 quasiparticles. It has recently been noted
 that the fractional charge which were not understood
 in the begining are due to electron clusters where 
spin $1/2$ is not sufficient. In these clusters the
 spin may be larger than $1/2$ such as $3/2$, 
$5/2$, $7/2$, etc. The repulsive Coulomb interactions
 align the 
electron spins ferromagnetically so that the spin of
 a cluster 
depends on the number of electrons[5]. Since usually 
a large 
magnetic field is present, the spins align parallel to
 the field although some may also be directed opposite
 to the magnetic field.
 So the electrons align even though there is no 
ferromagnetism. Our report of the new resonances makes 
use of the experimental
 measurements carried out by Syed et al[6].

\noindent{\bf 2. Theory}.\\

     The cyclotron resonance consists of a single
 resonance line at,
\beg
\omega_c={eB\over m^*c}.
\eeq
We predict new resonance lines at,
\beg
\omega_{c+}={1\over 2}g_+{eB\over m^*c}
\eeq
and at,
\beg
\omega_{c-}={1\over 2}g_-{eB\over m^*c}
\eeq
where
\beg
{1\over 2}g_{\pm} ={{\it l}+{1\over 2}\pm s\over
 2{\it l}+1}
\eeq
as described in ref.2. The spin is not restricted to 
s=1/2 
only. When there are electron clusters, it may be
 larger value 
also. The s=+1/2 is the Kramers time reversed state of 
s=-1/2. Therefore, ${1\over 2}g_{\pm}$ is having two 
values. When we 
reverse one spin from the $N_{\uparrow}$ state and put
 it in the $N_{\downarrow}$ state, the spin of the 
system changes by 2s. 
This is a text book problem which shows that,
\beg
{N_{\uparrow}-N_{\downarrow}\over 2}= s\,\,\,
proportional\,\,\, to \,\,N^{1/2}
\eeq
as given by Kittel and Kroemer[7] for ordinary Gaussian
 
distribution. Since, the energy in the centre of two 
Kramers 
conjugate states will
be proportional to $N_{\uparrow}-N_{\downarrow}=2s$, 
we
 expect
 that 
it varies as the square root of 2-dimensional electron 
density. 
Thus we have two new resonances at $\omega_{c+}$ and 
$\omega_{c-}$
with Kramers symmetry and Gaussian distribution for
 the central 
energy.

\noindent{\bf 2.~~Analysis of data}

     We will show that all of the above discussed
 properties can
 be extracted from the experimental work of Syed 
et al[6] and the 
new resonances at $\omega_{c\pm}$ can be identified
 from the data.
 The far infrared transmission data of a two-dimensional
 electron
 gas (2DEG) of density $1.14\times 10^{12}cm^{-2}$ in
 AlGaN/GaN at 
12.5 T shows a strong resonance at 6.9 meV and a weaker
 one to 5.1
 meV. We assign 6.9 meV to $g_+\mu_B12.5\times 10^4$ 
and 5.1 meV
 to $g_-\mu_B 12.5\times 10^4$. The value of $g_+$ 
is obtained as 
follows.
\beg
g_+9.274\times10^{-21}\times 12.5\times 
10^4=6.9\times10^{-3}
\times 1.602\times10^{-12}
\eeq
where the value of the Bohr magneton is
 $\mu_B=9.274\times10^{-21}$
erg/Gauss and the magnetic field is $12.5\times 
10^4$ Gauss.The resonance occurs at $6.9\times 
10^{-3}$ eV and we multiply it by $1.602\times
 10^{-12}$ to obtain erg units. This gives,
\beg
g_+=9.5353
\eeq
Similarly using the resonance at 5.1 meV, we obtain,
\beg
g_-=7.047
\eeq
From the above two values we obtain
\beg
{{1\over 2}g_+\over{1\over 2}g_++{1\over 2}g_-}=0.5749
\eeq
and
\beg
{{1\over 2}g_-\over {1\over 2}g_++{1\over 2}g_-}=
 0.4250.
\eeq
The sum of the above two numbers is 0.9999. According 
to one of 
our theorems $\nu_++\nu_-=1$. Therefore 0.9999 is just
 what we 
expected. Therefore the interpretation of resonances at
 6.9 meV 
and at 5.1 meV in terms of $g_+$ and $g_-$ is correct. 
Thus the new radiation at $\omega_{c+}$ and at
 $\omega_{c-}$ is discovered. It 
shows that the usual cyclotron resonance occuring at
 $\omega_c$ is flanked by two new resonances at
 $\omega_{c\pm}$. Some times, the prefactors may be 
zero or one, in which case $\omega_{c\pm}$ will 
occur in such a way that $\omega_c$ will not occur.

     The spin of $g_+$ is + and the spin of $g_-$ 
is -, so when one
spin is removed from $N_{\uparrow}$ and put in
 $N_{\downarrow}$, the 
spin excess is 2s. For Gaussian distribution, the
 centre of two new resonances varies as the square 
root of the number density of two-dimensional
 electrons. Indeed, the variation of this point is
 already plotted in ref.6 and the measured value 
agrees with the predicted 
square root of the number density.

     The resonance condition varies linearly with 
magnetic field. However, there is a crossing point 
or the centre of the two lines 
at $\omega_{c\pm}$. The energy levels at
${1\over2}g_-\omega_c
(n+{1\over 2})$ are narrowly spaced. When the magnetic 
field is increased, these narrowly spaced levels
 separate out untill their separation can become 
equal to those of ${1\over 2}g_+\omega_c
(n+{1\over 2})$. Thus there is a crossing point. 
The states are characterized by ${\it l}$ and $s$ and
 the Landau level number $n$. 
Thus the states are of the form $|n,{\it l}, 
\uparrow>$ and
 $|n,{\it l},\downarrow>$. Near the crossing point,
 the states get mixed so that the proper way of writing
 the wave function becomes,
\beg
{1\over\sqrt 2}(|n, {\it l},\uparrow> \pm |n', 
{\it l}',\downarrow>)
\eeq

This is the reason why energy as a function of magnetic
 field 
bends near the crossing point. The resonances above
 6 meV are $g_+$
type and below 6 meV are $g_-$ type. When energy is
 plotted as a function of magnetic field, bending 
occurs just as predicted. The prediction of new 
radiation at $\omega_{c\pm}$ is thus confirmed by 
the experiments.

\noindent{\bf3.~~ Conclusions}.

 We predict new resonances at $\omega_{c\pm}$. Their 
frequency locks 
the spin as given in the expressions. The centre of
 these lines 
varies as the square root of the number density of 
two-dimensional electron gas. The energy bends near
 the central point due to mixing 
of states. This represents fundamentally different 
resonance than the cyclotron resonance, known since 
1953, because the value of the spin
 enters in the frequency whereas the frequency of the 
cyclotron resonance is independent of the same. The
 cyclotron resonance is only
 one frequency, whereas ${\it many\,\,\, different
\,\,\, values}$ can be detected
 in the new resonances because of the many different 
values of 
${\it l}$ and $s$. Some of the details of the new
 resonance 
frequencies can be derived from the results given 
in ref.2.

\vskip1.25cm

\noindent{\bf4.~~References}
\begin{enumerate}
\item G. Dresselhaus, A. F. Kip and C. Kittel, Phys.
 Rev. {\bf 92}, 827 (1953).
\item K.N. Shrivastava, Introduction to quantum Hall
 effect,\\ 
      Nova Science Pub. Inc., N. Y. (2002).
\item K. N. Shrivastava, cond-mat/0212552.
\item K. N. Shrivastava, cond-mat/0303309, 
cond-mat/0303621.
\item K. N. Shrivastava, cond-mat/0302610.
\item S. Syed, M. J. Manfra, Y. J. Wang, H. L. Stormer
 and R. J. Molner, cond-mat/0305358.
\item C. Kittel and H. Kroemer, Thermal Physics, 
W. H. Freeman and Co., San Francisco, 1980, Second
 Edition, p.22.
\end{enumerate}
\vskip0.1cm

\end{document}